\documentclass[aps,prl,preprint,superscriptaddress,showpacs,showkeys]{revtex4-1}
\newcommand{\Real}{\mbox{Re}}

\usepackage{graphicx}
\usepackage{amssymb}
\usepackage{amsmath}
\usepackage{color}
\usepackage{bm}
\begin{document}

\title{Dielectrophoretic Equilibrium of Complex Particles}

\author{Tom Elkeles}
    \thanks{Both authors contributed equally to this work.}
    \affiliation{Faculty of Mechanical Engineering, Technion-Israel Institute of Technology, Israel.}

\author{Pablo Garc\'ia-S\'anchez}
    \thanks{Both authors contributed equally to this work.}
    \affiliation{Depto. Electr\'onica y Electromagnetismo. Facultad de F\'{\i}sica. Universidad de Sevilla. Avda. Reina Mercedes s/n, 41012. Sevilla, Spain.} 
    
\author{Wu Yue}
    \affiliation{Faculty of Mechanical Engineering, Technion-Israel Institute of Technology, Israel.}

\author{Antonio Ramos}
    \affiliation{Depto. Electr\'onica y Electromagnetismo. Facultad de F\'{\i}sica. Universidad de Sevilla. Avda. Reina Mercedes s/n, 41012. Sevilla, Spain.}

\author{Gilad Yossifon}
 \email{Corresponding author: yossifon@technion.ac.il}
 
 \affiliation{Faculty of Mechanical Engineering, Technion-Israel Institute of Technology, Israel.}


\begin{abstract}
In contrast to the commonly used spherical Janus particles, here we used engineered Janus particles that are fabricated using photolithography technique for precise control over their geometry and coated regions. Specifically, we studied a lollipop-shaped complex particle where its head is coated with gold while its tail is left bare. Due to their distinct electrical properties (i.e. electrical polarizability) the particle exhibits force equilibrium where opposite dielectrophoretic forces acting on its head and tail exactly cancel each other to yield a stable equilibrium position. This was realized in a quadrupolar electrode array where the equilibrium position of the engineered particle could be tuned by the frequency. This stands in contrast to the standard dielectrophoretic behavior where the particle shifts positions from either the center of the quad to the very edge of the electrodes when shifting from a negative to positive dielectrophoretic response, respectively. This opens new opportunities for positioning control of such complex particles for self-assembly, biosensing, biomimetic spermatozoa and more.
\end{abstract}

\maketitle
Dielectrophoresis (DEP), defined as the translational motion of neutral particles due to the effects of polarization in a nonuniform electric field, is a well-established technique that exploits their unique dielectric properties to manipulate particles under alternating current (AC) fields \cite{jones01}. In particular, a crossover frequency (COF), which is the AC frequency at which the DEP force vanishes, i.e. particles shift from attraction (positive DEP, pDEP) to repulsion (negative DEP, nDEP), can be used as a sensitive discriminator between different particle types or conditions \cite{ben-bassat15,zhang15}. The time-averaged DEP force for a homogeneous dielectric spherical particle suspended within an electrolyte under a non-uniform electric field is represented by $\bm{F}_{\mathrm{DEP}}=\pi\varepsilon_e R^3\Real[\tilde{K}]\nabla |E|^2$, where $R$ is the radius of the particle, $E$ is the amplitude of electric field, $\varepsilon_e$ is the permittivity of the electrolyte and $\Real[\tilde{K}]$ is the real part of the Clausius-Mossotti (CM) factor \cite{morgan03,gagnon11,jones03}. The CM factor is defined as $\tilde{K}(\omega)=(\tilde{\varepsilon_p}-\tilde{\varepsilon_e})/(\tilde{\varepsilon_p}+2\tilde{\varepsilon_e})$, $\tilde{\varepsilon}=\varepsilon+\sigma/(i\omega)$ , where $\tilde{\varepsilon_p}$ and $\tilde{\varepsilon_e}$ are the complex permittivities of the particle and the electrolyte, respectively, and $\varepsilon$ and $\sigma$ represent the real permittivity and the conductivity, respectively. The CM depends on the frequency of the electric field, which determines both the direction of the DEP force and its magnitude. The CM factor of micron size dielectric particles exhibits mostly nDEP, except dielectric nanoparticles at relatively low conductivity solutions wherein their surface conductance dominates \cite{okonski60}. In contrast, the CM factor of metallic coated particles has been shown to switch sign from nDEP to pDEP with increasing frequency as the induced electrical double layer (EDL) formed on the metallic coating shifts from full electrical screening to no-screening at frequencies sufficiently larger than the RC frequency \cite{Garcia-Sanchez12,arcenegui13b,arcenegui14}.\\

\begin{figure}[h!]
    \centering
    \includegraphics[width=\columnwidth]{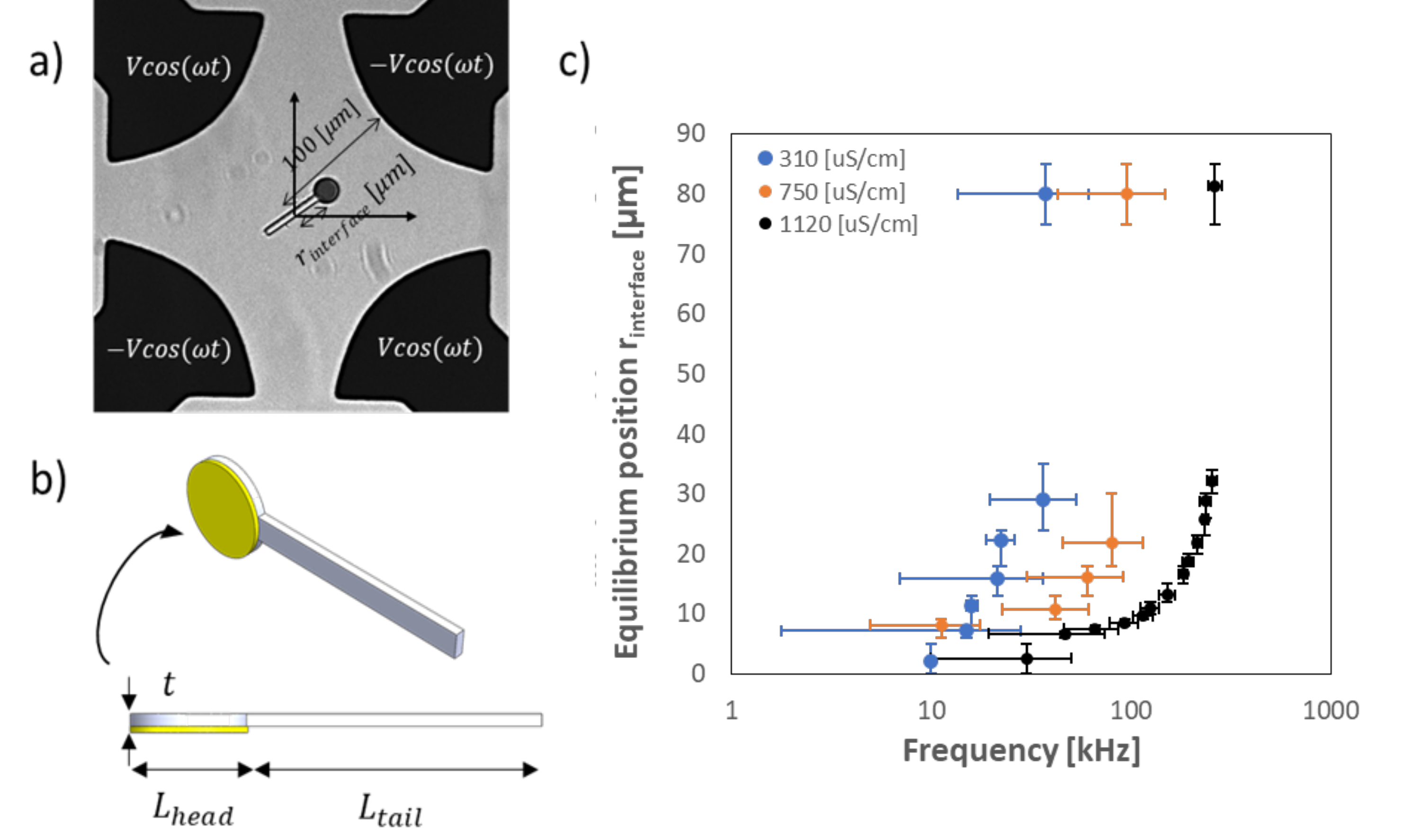}
    \caption{a) Microscope image of the complex particle within a quadrupolar electrode array. Here we define the equilibrium position $r_{\mathrm{interface}}$ from the center of the quad to the interface of the tail and the head of the particle; b) Schematics of the complex particle with dimensions $L_{\mathrm{tail}}=50\,\mathrm{\mu m}$, $L_{\mathrm{head}}=20\,\mathrm{\mu m}$, $t=2\,\mathrm{\mu m}$ while the thickness of the gold coating is of 30\,nm on top of a 10\,nm Cr coating (marked with yellow in the scheme); c) Equilibrium position of the hybrid particle as a function of the electric field frequency. The graph depicts three different electrolyte conductivities at voltage of either 15\,V or 20\,V with no observable difference in the equilibrium position for these two voltages. Error bars represent the standard deviation of several (3-5) experiments. See supplemental movie 1.}
    \label{figure1}
\end{figure}

Herein, we will exploit this transition of the polarizability of the metallic coating of a complex particle in order to control the interplay between its different coated and non-coated (i.e. dielectric) parts. Such an interplay between the coated and non-coated parts of a particle has been demonstrated for a spherical metallo-dielectric Janus particle (JP) where its overall DEP and electrorotation (ROT) behavior was taken to be an average of the same spherical particle as though it was uniformly dielectric and uniformly coated \cite{chen16}. However, as noted for a certain frequency the DEP response of the JP was either nDEP or pDEP. Hence, no intermediate equilibrium positions were obtained within the quadrupolar electrode array. In contrast, here we will demonstrate that if we engineer the Janus particle to have a large enough separation between its dielectric and metallic coated parts similar to sperm cell \cite{shuchat2019}, such intermediate equilibrium positions could be achieved. Such engineered Janus particle with controlled geometry and selective metallic coating is obtained using standard photolithography fabrication technique \cite{shields18} (see supplementary materials for details of the particle fabrication).\\

Janus particles in general is an emerging field of research that attracted immense attention in the recent years due to their behavior as a self-propelling (active) particles. These active particles found applications in a broad range of areas such as drug delivery, detoxification, environmental remediation, immunosensing, etc \cite{kherzi16}. As opposed to phoretically driven transport, which is characterized by mass migration in response to externally imposed gradients, active particles asymmetrically draw and dissipate energy at the colloidal scale, creating local gradients which drive autonomous propulsion \cite{han18}. Since the driving force is produced on the particle level, active colloids are free to travel along individual pathlines. In particular, metallo-dielectric JP represent a unique subset of active colloids, where energy source is an externally applied electric field where the variation of the frequency of applied electric field has been shown to alter both the speed and direction \cite{yan16,boymelgreen16,squires06}.\\

\begin{figure}
    \centering
    \includegraphics[width=\columnwidth]{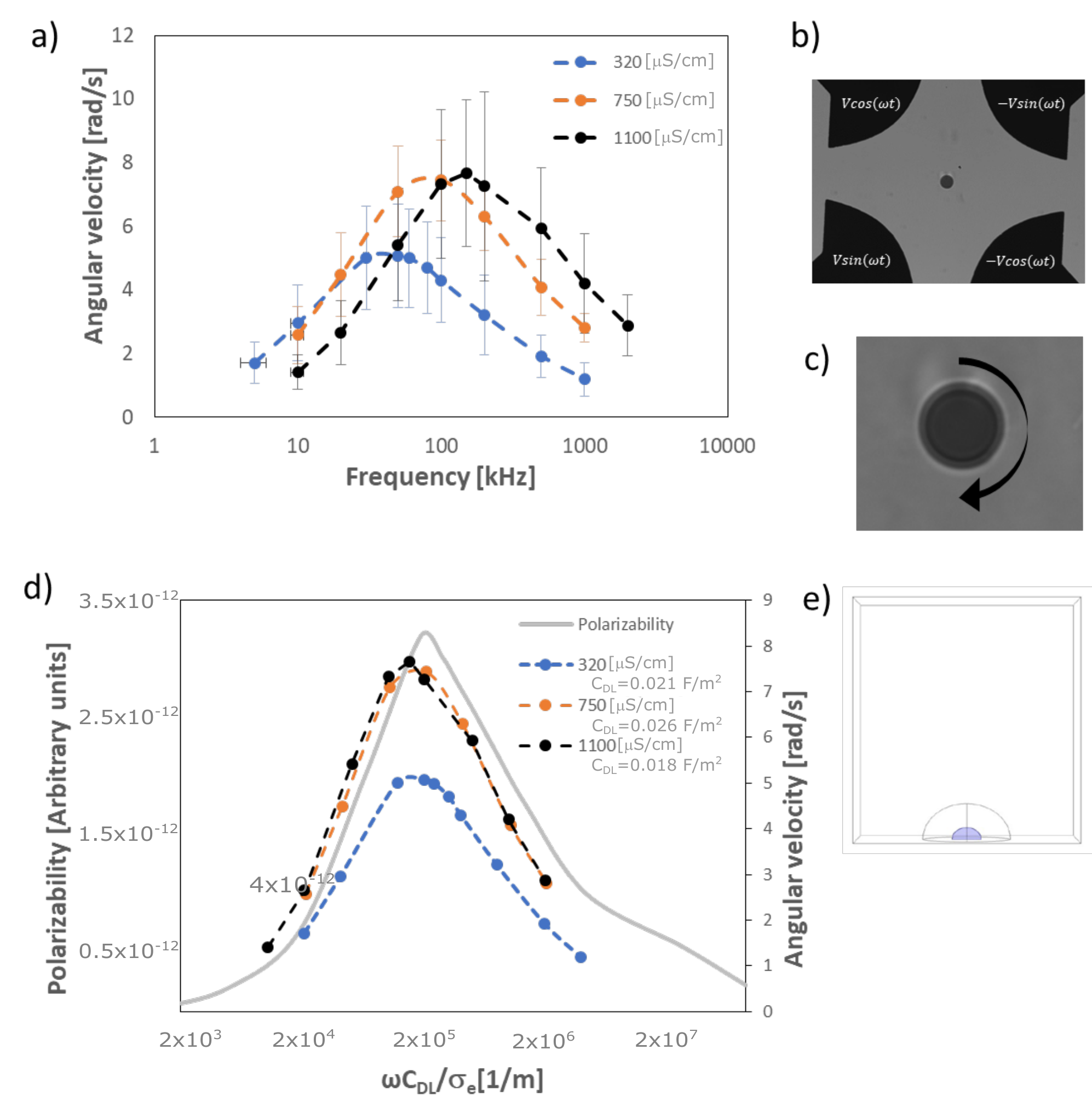}
    \caption{a) ROT velocity of the Au coated disk (only the `head' part of the complex particle) as a function of the frequency at a voltage of 20\,V. Error bars represent the standard deviation of 4-5 experiments; b-c) Microscope image of the coated disk and its magnified image along with the direction of rotation as observed from the inverted microscope. See supplemental movie 2; d) ROT velocity of a coated disk as a function of the normalized frequency. The graph shows also the polarizability that was calculated numerically; the extracted $C_{\mathrm{DL}}$ values are indicated in the legend; e) The computational domain of the numerical simulation. 
}
    \label{figure2}
\end{figure}

\begin{figure}
    \centering
    \includegraphics[width=\columnwidth]{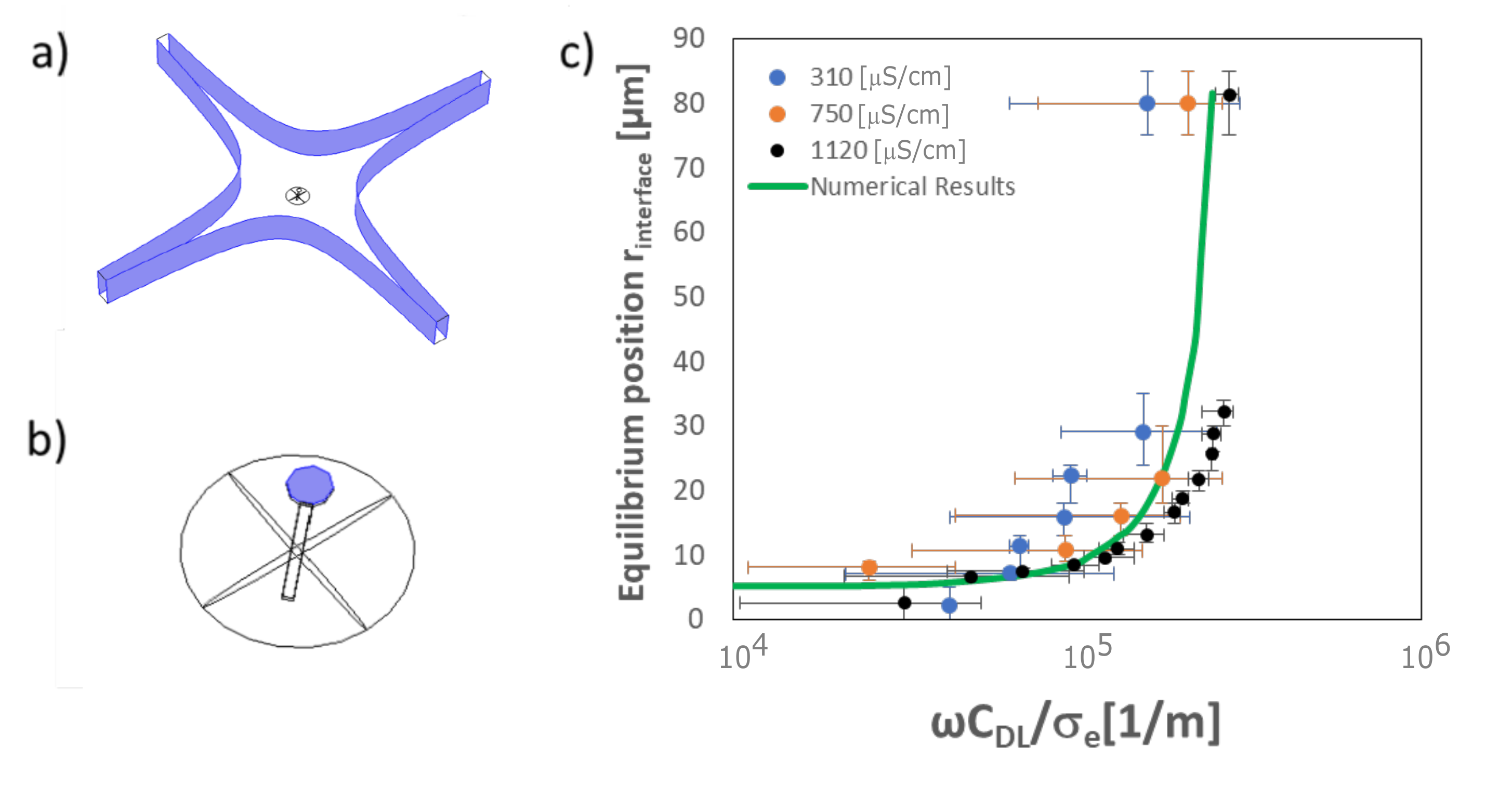}
    \caption{a) The computational domain of the numerical simulation showing the quadrupolar electrodes as well as the complex particle; b) magnified image of the complex particle enveloped within an ellipsoidal surface at which the Maxwell-stress integration was performed in determining the DEP force; c) Experimental and numerical results of the particle position as a function of the normalized frequency.}
    \label{figure3}
\end{figure}

\begin{figure} [h!]
    \centering
    \includegraphics[width=\columnwidth]{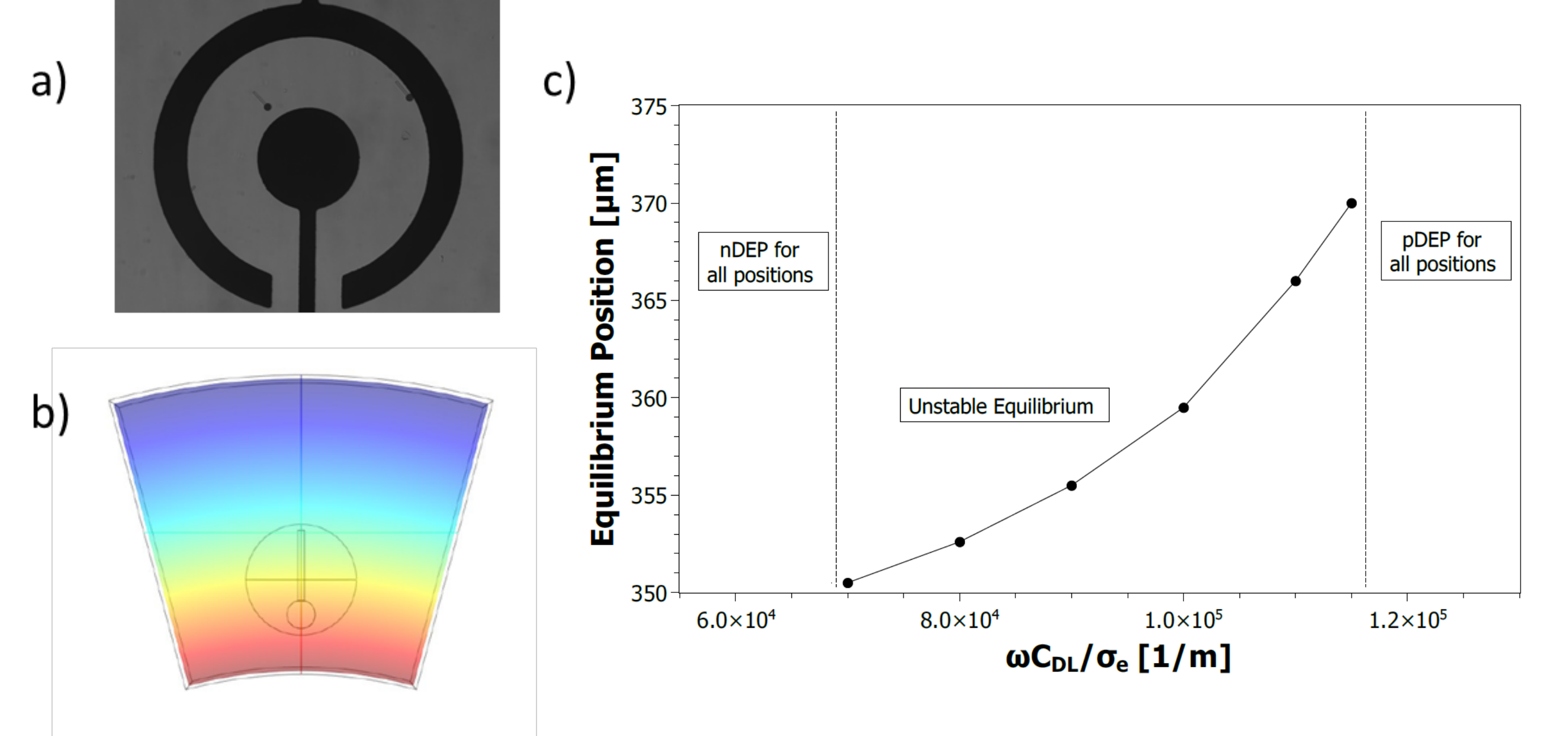}
    \caption{a) Microscope image of the complex particle inside the ring-shaped electrode array with a gap of $200\,\mathrm{\mu m}$ gap between the inner and outer electrodes; b) Computational domain of the numerical simulation with the particle enveloped within a spheroidal surface; c) Numerical simulation results indicating that there is an unstable equilibrium in the frequency regime between a nDEP and pDEP response. See supplemental movie 4.}
    \label{figure4}
\end{figure}

The chips are fabricated using standard photolithography and consist of either a quadrupolar with a gap of 200\,$\mathrm{\mu m}$ and 500\,$\mathrm{\mu m}$ for DEP (Fig. \ref{figure1}) and ROT (Fig. \ref{figure2}) characterization, respectively, as well as a ring electrode array with a gap of 200\,$\mathrm{\mu m}$ between the electrodes for control purposes acting as an electrode geometry where no stable equilibrium can exist (Fig. \ref{figure4}). The electrolyte solution consists of dissolved KCl salt within DI water at different concentrations which is then introduced into a shallow chamber (height of 120\,$\mathrm{\mu m}$) covering the electrode array along with the `lollipop' particles. In order to reduce adsorption, the chamber is pretreated with 0.2\%\,v/v of Tween 20 (Sigma). The electrodes are connected to a function generator, which we use to control the frequency and amplitude of the applied electric field. For the ROT setup we used a rotating electric field such that is has 90 degrees phase shift between the electrodes.
A complex particle can be potentially stagnant under the following modes: (1) at the center of the quad under nDEP response \cite{voldman2001}; (2) at stable equilibrium positions within the quad where the two counteracting DEP forces exactly cancel each other; (3) at the edge of the electrode under pDEP response. This has been verified for the `lollipop' shaped particle where the DEP equilibrium (i.e. mode 2) positions were varied with frequency. Figure 1a shows such complex particle in stable equilibrium mode, when the frequency of the alternating current electric field is such that parts 1 and 2 of the particles (as defined in figure \ref{figure1}b) are experiencing DEP forces of comparable magnitude and opposite directions. In order to measure the stable equilibrium position of the particles, they were placed inside the quad while experiencing an electric field at a specific frequency. Figure \ref{figure1}c shows the average equilibrium position of 3-5 particles versus the frequency of the applied field. In order to assure that this is indeed a stable equilibrium position the particles were often shaken after which they have recovered their equilibrium position. In addition, frequencies were applied in different order to negate the existence of system hysteresis (see supplemental movie 1).\\

It was found that increasing the conductivity of the medium results in a wider range of these equilibrium frequency windows. In addition, we found that the frequencies at which the particle transitions from nDEP to stable equilibrium and from stable equilibrium to pDEP strongly depended on the solution conductivity and shifts to higher values with increasing solution conductivity. Whereas the dielectric (SU8) tail of the particle always exhibit nDEP behavior the metallic coated head of the particle switches from nDEP to pDEP behavior with increasing frequency at a transition frequency that corresponds to the RC time of the induced EDL. This can explain the shift of the above transitions observed for the complex particle as the RC frequency increases linearly with $\sigma_e$.
\\

In order to match the experimental results to the numerical results with the angular frequency normalized by the conductivity and the specific capacitance of the electric double layer, $C_{\mathrm{DL}}$, on the metallic coated disk we conducted several experiments of electro-rotating field (ROT) from which we extracted the $C_{\mathrm{DL}}$. Due to the relatively high voltage applied on the particle, exceeding the thermal potential, the linear Debye-Huckel model for the capacitance is not applicable. Furthermore, there is always the ambiguity regarding the Stern layer capacitance which needs to be somehow resolved. Therefore, the value of the $C_{\mathrm{DL}}$ was experimentally extracted for each solution conductivity only for the coated disk particle without the tail section. A rotating electric field was applied in the 500\,$\mathrm{\mu m}$ quadrupolar electrode array and the angular velocity of the gold coated disk was measured in different frequencies \cite{park17}. After obtaining the ROT spectra for each conductivity (Fig. \ref{figure2}a) we used the $C_{\mathrm{DL}}$ as a fitting parameter in the normalized experimental curves to match the normalized angular velocities of the point of maximum angular velocity to that obtained in the numerical simulation (Fig. \ref{figure2}d). Following the extraction of the $C_{\mathrm{DL}}$ values the experimental results of the equilibrium position were then depicted using a normalized angular frequency and showed qualitative agreement with the numerical simulations (Fig. \ref{figure3}). For simplification, we have neglected the contribution of electro-convection on this electrostatic force balance as it is expected to be suppressed due to the shallowness of the microchamber height (120\,$\mathrm{\mu m}$) (see supplemental movie 3).\\

For a stable equilibrium to occur the non-uniform electric field must obey a certain spatial distribution (see supplemental figure S1). While the quadrupolar electrode array does satisfy the necessary conditions other electrode arrays such as the ring electrode shown in Fig. \ref{figure4} does not. In accordance to the heuristic model described in the supplemental materials no such stable equilibrium was experimentally found in such a ring electrode setup. Instead, it has been found that the particles were either repelled from the inner electrode or attracted to it. The results of the simulations suggested that in the ring geometry exists a range of frequencies that the particle experiences an unstable equilibrium (Fig. \ref{figure4}c). Depending on the initial radial distance of the particle from the center of the array the response could be either pDEP or nDEP if it is smaller (i.e. $r_{\mathrm{initial}}<r^{\mathrm{eq}}$) or larger (i.e. $r_{\mathrm{initial}}>r^{\mathrm{eq}}$) than the equilibrium position, respectively (Fig. S1). Hence, due to the unstable equilibrium there are eventually two terminal positions of the particle, either at the inner electrode edge due to the dominance of pDEP or close to the edge of the outer ring electrode due to the dominance of nDEP. It is worth noting that since our system is not an ideal 2D geometry but is actually 3D, and hence there is an electric field intensification at the very edge of the outer ring electrode --- the potential well of the nDEP is somewhat away from the edge of the outer ring electrode (see supplemental movie 4).\\

We have demonstrated that a complex engineered particle, where opposite DEP forces can occur at its different parts, exhibits a stable equilibrium within a frequency range. In addition, we found that the frequencies at which the particle transitions from nDEP to stable equilibrium and from stable equilibrium to pDEP strongly depended on the solution conductivity as expected. A quantitative comparison between the numerical results of the equilibrium position as function of frequency to the experimental results validates it and gives further information to the dependency of the equilibrium range with the solution conductivity. ROT spectra on the coated disk shows that indeed the metallic part of the particle undergoes transition from nDEP to pDEP as expected and the frequency of maximum angular velocity increases with increasing conductivity. This ability to control the equilibrium position of complex particles by simply changing the applied electric field frequency may open new opportunities for self-assembly, collective behavior and biosensing.\\

\section*{Acknowledgments}
We wish to acknowledge the Technion Russel-Berrie Nanotechnology Institute (RBNI) and the Technion Micro-Nano Fabrication Unit (MNFU) for their technical support. PGS and AR acknowledge financial support by ERDF and Spanish Research Agency MCI under contract PGC2018-099217-B-I00.

\end{document}